\documentclass[aps, prl, twocolumn, floatfix, groupaddress, reprint, 10pt]{revtex4-1}  
\usepackage{bm}
\usepackage{graphicx}
\usepackage[usenames]{color}
\usepackage{microtype}
\usepackage{amsmath}
\usepackage{amssymb}
\usepackage{xspace}
\usepackage{footnote}
\usepackage{relsize}
\usepackage{hhline}

\usepackage{physics}

\newcommand{\iu}{\mathrm{i}}

\newcommand{\br}{\mathbf{r}}

\newcommand{\bk}{\mathbf{k}}

\newcommand{\dbr}{\textrm{d}\br}

\newcommand{\brhat}{\hat{\mathbf{r}}}

\newcommand{\CA}{\mathcal{A}}
\newcommand{\CV}{\mathcal{V}}
\newcommand{\CVp}{\mathcal{V}^p}

\newcommand{\ibraket}[3]{\braket{#1}{#2}_{#3}}

\begin{document}

\title{
Incorporating long-range physics in atomic-scale machine learning
}

\author{Andrea Grisafi}
\author{Michele Ceriotti}

\affiliation{Laboratory of Computational Science and Modeling, IMX, \'Ecole Polytechnique F\'ed\'erale de Lausanne, 1015 Lausanne, Switzerland}

\begin{abstract}
The most successful and popular machine learning models of atomic-scale properties  derive their transferability from a locality ansatz. 
The properties of a large molecule or a bulk material are written as a sum over contributions that depend on the configurations within finite atom-centered environments. 
The obvious downside of this approach is that it cannot capture non-local, non-additive effects such as those arising due to long-range electrostatics or quantum interference. 
We propose a solution to this problem by introducing non-local representations of the system that are remapped as feature vectors that are defined locally and are equivariant in $O$(3). We consider in particular one form that has the same asymptotic behavior as the electrostatic potential. We demonstrate that this framework can capture non-local, long-range physics by building a model for the electrostatic energy of randomly distributed point-charges, for the unrelaxed binding curves of charged organic molecular dimers, and for the electronic dielectric response of liquid water.
By combining a representation of the system that is sensitive to long-range correlations with the transferability of an atom-centered additive model, this method outperforms current state-of-the-art machine-learning schemes, and provides a conceptual framework to incorporate non-local physics into atomistic machine learning.
\end{abstract}

\maketitle

\section{Introduction}

In recent years, atomistic machine learning models have become increasingly popular as a way to perform fast predictions of molecular and material properties with the accuracy of first-principle quantum mechanical calculations~\cite{behler2017}, but a much reduced cost.
The success of these methods has gone hand-in-hand with the progress in constructing representations for molecular and materials configurations that are flexible enough to be transferred across a wide spectrum of different atomic arrangements, while satisfying, at the same time, stringent symmetry constraints~\cite{behler2007,bartok2013,Shapeev2015,Glielmo2017,grisafi2018}. 

At the core of the vast majority of transferable machine-learning model for physical properties lies the local nature of the underlying atomistic representation. 
This is usually constructed by considering the set of atomic coordinates that are included within spherical environments of a given radial cutoff around any arbitrary atomic center~\cite{bartok2017,Chmiela2017,Zhang2018}. 
The prediction of a given physical property is therefore formally decomposed in the sum of atom-centered contributions that effectively incorporate information associated with many-body structural correlations between atoms in each local environment. This locality assumption is very convenient, as it keeps at bay the dimensionality of the regression problem that is modelled by ML, and is physically justified by the nearsigthedness principle of electronic matter~\cite{Prodan2005}.  
The major drawback is that  it neglects long-range physical effects. Long-range electrostatic interactions, for example, are known to play a fundamental role in the description of ionic systems~\cite{kjellander2018}, macroscopically polarized interfaces~\cite{guo2018}, electrode surfaces~\cite{jorn2013} and nano-science in general~\cite{french2010}.
In all these cases, the pathologically slow decay $\sim 1/r$ of the Coulomb interaction makes it virtually impossible to reach convergence while using a local machine-learning scheme, which is reflected in an effective limit to the accuracy that can be reached by these models. 

The problem of incorporating long-range effects in electronic energy predictions is usually tackled by explicitly separating the local many-body contribution to the total energy from a classical electrostatic term approximated via pairwise Coulomb interactions. 
This can be done either by direct subtraction of the Ewald-like electrostatic energy of the system~\cite{bartok2010,Deng2019}, or by machine learning, in turn, the partial charges and the atomic multipoles that determine the long-range electrostatics~\cite{Artrith2011,Bereau2015,Bereau2017,Bleiziffer2018,Nebgen2018,Yao2018}.
Other more sophisticated models, specifically designed for ionic systems, rely on a charge equilibration scheme~\cite{Ghasemi2015,Faraji2017}. %

Beyond electrostatic energies, the breakdown of a local machine learning model is particularly pronounced when dealing with intrinsically non-local quantities like the dielectric response of a condensed-phase medium~\cite{grisafi2018}. This non-locality has to do both with the effect of the far-field electrostatics~\cite{bottcher1978}, and to the topological quantum nature of the macroscopic polarization of an infinitely extended material~\cite{resta1994,Resta2010}. 
In this case, the problem can possibly be bypassed by adopting specific physical prescriptions. Examples of this can be found in Ref.~\cite{grisafi2018}, where the dielectric tensor $\boldsymbol{\varepsilon}_\infty$ of liquid water is learned indirectly by building a model for an effective molecular polarizability that is mapped to $\boldsymbol{\varepsilon}_\infty$ through the Clausius-Mossotti relationship~\cite{bottcher1978}. 
In the context of reproducing the autocorrelation function of the macroscopic polarization of liquid water, another strategy has recently been adopted, where the selected learning targets are the positions of the Wannier centers that are used to recast the electron density of the system into a set of point-charges~\cite{Zhang2019arxiv}.

By and large, the learning models previously described tackle the problem of including long-range phenomena by making use of an \textit{ad hoc} definition of the electrostatic energy, or dielectric response, in terms of local atomic quantities. 
Although successful, these kind of approaches have the downside of being very system dependent and, as such, hardly transferable across systems that have a different nature, e.g., those related to charge transfer, or to charge polarizability in (near)-metallic systems~\cite{Wilkins2019}. 
Capturing long-range effects without any prior assumption on the nature of the learning target is a difficult task to accomplish with the methods currently available. Most of the approaches that have explicitly attempted to do so, such as Coulomb kernels~\cite{Rupp2012}, many-body tensor representations~\cite{Huo2017arxiv}, or multi-scale invariants~\cite{Hirn2017}, are built upon a global representation of the system rather than on an additive atom-centred model. 

Here we propose a simple, yet elegant, solution to this problem, where the non-local character of the target property is incorporated in a symmetry-equivariant fashion into an atom-centered representation. In doing so, we construct a formalism that ensures that the resulting features exhibit the correct asymptotic dependence on the distribution of atoms in the far-field. 
This representation can be incorporated straightforwardly into conventional, additive machine learning models.
While the idea is very general, we present as an example a model that has an asymptotic behavior consistent with electrostatic interactions. We show that it can be used successfully to build a local machine learning model that accurately reproduces Coulomb interactions between point particles, the binding curves of charged organic fragments, and the electronic dielectric response of bulk water. 

\section{Long-distance equivariant representation}

Let us start from the same formal definition of a ML representation of a structure $\CA$ that was introduced in Ref.~\cite{willatt2019}, written in the position basis as a decorated atom density
\begin{equation}
\bra{\br}\ket{\CA} = \sum_i g(\br -\br_i) \ket{\alpha_i},
\label{eq:rA-ket}
\end{equation}
where the index $i$ runs over all the atoms in the structure, $g$ is a Gaussian (or another localized function) peaked at each atom's position $\br_i$, and $\ket{\alpha_i}$ is an abstract vector that encodes the chemical nature of the atom.
We now introduce an atom-density potential representation 
\begin{equation}
\bra{\br}\ket{\CVp} = \sum_i  \ket{\alpha_i} \int \dbr'\, \frac{g(\br' -\br_i)}{\left|\br'-\br\right|^p}.  \label{eq:rVp-ket}
\end{equation}
The rationale for performing this transformation (that can be seen as the action of a linear integral operator on $\ket{\CA}$) is that, whereas $\bra{\br}\ket{\CA}$ contains information only about the atoms in the vicinity of $\br$, $\bra{\br}\ket{\CVp}$ contains information about the position of \emph{all} atoms in the structure, with a dependence on the position of the $i$-th atom that decays asymptotically as $\left|\br-\br_i\right|^{-p}$\footnote{Evaluation of the integral for $p>1$ require some form of regularization or short-distance cutoff to remove the singularity for $\br\rightarrow\br_i$}. 
The physical significance of $\ket{\CVp}$ is obvious, if one considers typical forms of the interactions between atoms and molecules.
For instance, if we had a single species and interpreted \eqref{eq:rA-ket} as a charge density, $\bra{\br}\ket{\CV^{1}}$ would correspond to the electrostatic potential generated by such charge density. Analogously, the $p=6$ case would provide the formally correct asymptotic limit of the energy per particle associated with dispersion interactions~\cite{Dreizler2012}, which has inspired previous representations of local environmets such as aSLATM~\cite{Huang2019arxiv}. 

Proceeding as in Ref.~\citenum{willatt2019}, we can symmetrize the representation over the continuous translation group, taking a tensor product with the density representation to preserve structural information. One obtains the symmetrized ket
\begin{equation}
\ibraket{\br}{\CA\CVp}{\hat{t}} =
\int \dd \hat{t} \bra{\boldsymbol{0}}\hat{t}\ket{\CA}\bra{\br}\hat{t}\ket{\CVp} =
\sum_j  \ket{\alpha_j}\bra{\br}\ket{\CVp_j},
\label{eq:rAVp-ket}
\end{equation}
where we introduced the shorthand notation (see the SI for a full derivation)
\begin{equation}
\bra{\br}\ket{\CVp_j} = 
\sum_{i} \ket{\alpha_i} \int \dbr' \frac{(g\star g)(\br' -(\br_i-\br_j))}{\left|\br'-\br\right|^p}\label{eq:rVj-ket}.
\end{equation}
Modulo the re-definition of the atom density function as the auto-correlation of $g$, $\bra{\br}\ket{\CVp_j}$ is just the atom-density potential~\eqref{eq:rVp-ket} computed using $\br_j$ as the origin of the reference frame.

Symmetrization over the translation group leads naturally to a structural representation that amounts to a sum over atom-centred descriptors -- foreshadowing an additive property model built on such feature vector. Particularly for low values of the potential exponent $p$, however, the integral in Eq.~\eqref{eq:rVj-ket} introduces a substantially non-local behavior. The value of $\bra{\br}\ket{\CVp_j}$ in the vicinity of the central atom $j$ can in principle depend on the position of atoms that are very far from it, \emph{even if one introduces a cutoff function that restricts the range of $\bra{\br}\ket{\CVp_j}$ around the central atom, and hence its complexity}. 
One can then symmetrize further over the rotation group and over inversion symmetry. We will refer from now on to the resulting class of atomistic representations that capture long-range interactions based on the local value of an atom-density potential as the \textit{long-distance equivariant}~(LODE) framework. In the following we will focus on the case of $p=1$, that corresponds to electrostatic interactions.

It is instructive to first consider the case of the first order spherical invariant, and to take the limit in which the atom density is represented by Dirac-$\delta$ distributions. It is easy to see that in this limit
\begin{equation}
\bra{\alpha\br}\ket{\CV_j^1} = 
\sum_{i\in\alpha} \frac{1}{\left|\br - \br_{ij}\right|}\label{eq:rV1-delta},
\end{equation}
where $\br_{ij}=\br_i-\br_j$.
Integrating over the SO(3) group yields the first invariant
\begin{equation}
\bra{\alpha r}\ket{{\CV_j^1}^{(1)} } = 
\int \dd \hat{R} \bra{\alpha r\brhat}\hat{R}\ket{\CV^1_j}=
\sum_{i\in\alpha}   \min\left[\frac{1}{r},\frac{1}{r_{ij}}\right]\label{eq:rV1-delta-1},
\end{equation}
that simply sums up $1/r_{ij}$ terms for all atoms \emph{outside} the region over which the LODE representation is computed.
Ignoring the contribution from the atoms within the cutoff, that can be better characterized by other atomic structure representations, a linear model built on these features is equivalent to a fixed point-charge electrostatic model.
In other words, in this limit the radial dependence of the regression weights is integrated out, and the weights associated with each pair of central atom type $\alpha'$ and neighbor type $\alpha$ corresponds to the product of the atomic charges $q_{\alpha'}$ and $q_{\alpha}$.

While this construction is very revealing, it is clear that its descriptive power is limited. Non-linear kernel models can provide a more flexible functional form, but higher-order invariants provide a systematic way of incorporating more information on structural features. 
As in the SOAP framework for the atom density~\cite{bartok2013,de+16pccp,willatt2019}, the most convenient way to compute such invariants involves writing the scalar field associated with the species $\alpha$ on a basis of radial functions $R_n(r)$ and spherical harmonics~$Y^l_m(\brhat)$,
\begin{equation}
\bra{\alpha n l m}\ket{\CVp_j} = \int \dd \br\,  R_n(r) Y^{l}_{m}(\brhat)^\star \bra{\alpha \br}\ket{\CVp_j}
\label{eq:anlm-ket}
\end{equation}
and then computing the appropriate spherically-covariant combinations. For example, for rotationally invariant representations of order $\nu=2$ (the form that is equivalent to the SOAP power spectrum and that we will use in applications)
\begin{equation}
\bra{\alpha n \alpha' n' l}\ket{{\CVp_j}^{(2)}} = \sum_{|m|\le l} \frac{\bra{\alpha n l m}\ket{\CVp_j}^\star\bra{\alpha' n' l m}\ket{\CVp_j}}{\sqrt{2l+1}}
\label{eq:nnm-ket}.
\end{equation}
The extension to higher orders in spatial correlations $\nu>2$ and/or to rotationally covariant representations of a given spherical-tensor order $\lambda>0$ is straightforward based on the analogous density-based counterparts~\cite{grisafi2018,willatt2019,grisafi2019-arxiv}. 
Note that it is also possible to compute representations that combine different values of $p$, and even $p=0$, corresponding to the atom-density field. A systematic investigations of the various combinations, and their physical meaning, is left for future work. 

\subsection{Efficient evaluation of the LODE representation} 

As discussed in the SI, for molecules and clusters the expansion~\eqref{eq:anlm-ket} can be computed conveniently in real space, by numerical integration on appropriate atom-centred grids. For a bulk system, described by a periodically-repeated supercell, the long-range nature of the integral kernel that appears in~\eqref{eq:rVj-ket} would make computing the expansion prohibitive.
This is exactly the same problem one faces when evaluating electrostatic interactions in the condensed phase, and fortunately it has long been solved, e.g., with the many techniques based on the use of a plane-waves auxiliary basis~\cite{Ewald1921, Essmann1995}.
Consider the plane-wave definition as $\bra{\br}\ket{\bk} = e^{\iu\bk\cdot\br}$, with $\left\{\bk\right\}$ representing a set of wave-vectors that are compatible with the simulation box. The fact we start from a smooth, Gaussian atom density, means that in practice one needs only a manageable  number of plane waves. In particular, the width $\sigma$ of the Gaussian density determines the minimum wavelength that should be introduced in the the plane-wave expansion, so that $\bk$-vectors only need to be generated within a sphere of radius $k_\text{max}$ of the order of $2\pi/\sigma$. In order to evaluate the local potential projections,
it is then enough to include the identity resolution $\sum_{\bk}\ket{\bk}\bra{\bk}$ within the braket of Eq.~\eqref{eq:anlm-ket}, i.e.
\begin{equation}\label{eq:k-resolution}
\bra{\alpha n l m}\ket{\CVp_j}
     = \sum_{\bk}\bra{n l m}\ket{\bk}\bra{\alpha \bk}\ket{\CVp_j} .
\end{equation}
As detailed in the SI, $\bra{n l m}\ket{\bk}$ corresponds to the expansion in plane waves of the basis of the local environment representation, and can be computed analytically once and for all if the radial functions are taken to be Gaussian type orbitals~\cite{Cahill2013}. Conversely, $\bra{\alpha \bk}\ket{\CVp_j}$ represents the Fourier components of the potential generated by the Gaussian density of element $\alpha$ for the entire system, and can be readily computed analytically~\cite{Allen1989}.
As a result, the geometric local nature of the representation of Eq.~\eqref{eq:k-resolution} is formally factorized from its system-dependent global character. It has not escaped our attention that Eqn.~\eqref{eq:k-resolution} could also be used to compute efficiently the coefficients of the density expansion that enter, for instance, the SOAP framework.
In the context of electrostatic interactions, one should note that although the fictitious charge density distribution of Eq.~\eqref{eq:rA-ket} does not satisfy charge neutrality, one can avoid a divergence of the potential by ignoring the $\bk$=$\boldsymbol{0}$ component from the sum of Eq.~\eqref{eq:k-resolution}.
Similarly, divergences in the potential for $p>1$ can be eliminated by appropriately regularizing the $1/r^p$ divergence in reciprocal space. 

\section{Results}

We now proceed to test the performance of the LODE representation in the context of predicting scalar electrostatic properties. 
In all cases we use Gaussian process regression using simple polynomial kernels, to emphasize the role of the features - as opposed to the regression scheme - on the performance of the model. Details of the parameters used in each example are reported in the SI. 
We use the SOAP framework as the baseline for a comparison, which is appropriate given the close relation between the two approaches, and the excellent performances demonstrated by SOAP-based models.
It is important however to stress that \emph{any} local model with a finite cutoff will exhibit similar behavior as what we observe with SOAP.
We also benchmark the combination of SOAP and LODE, that incorporates the advantages of both short-range and long-range models, realizing a kind of range-separated machine learning framework. 

\begin{figure}[tbp]
    \centering
    \includegraphics[width=9cm]{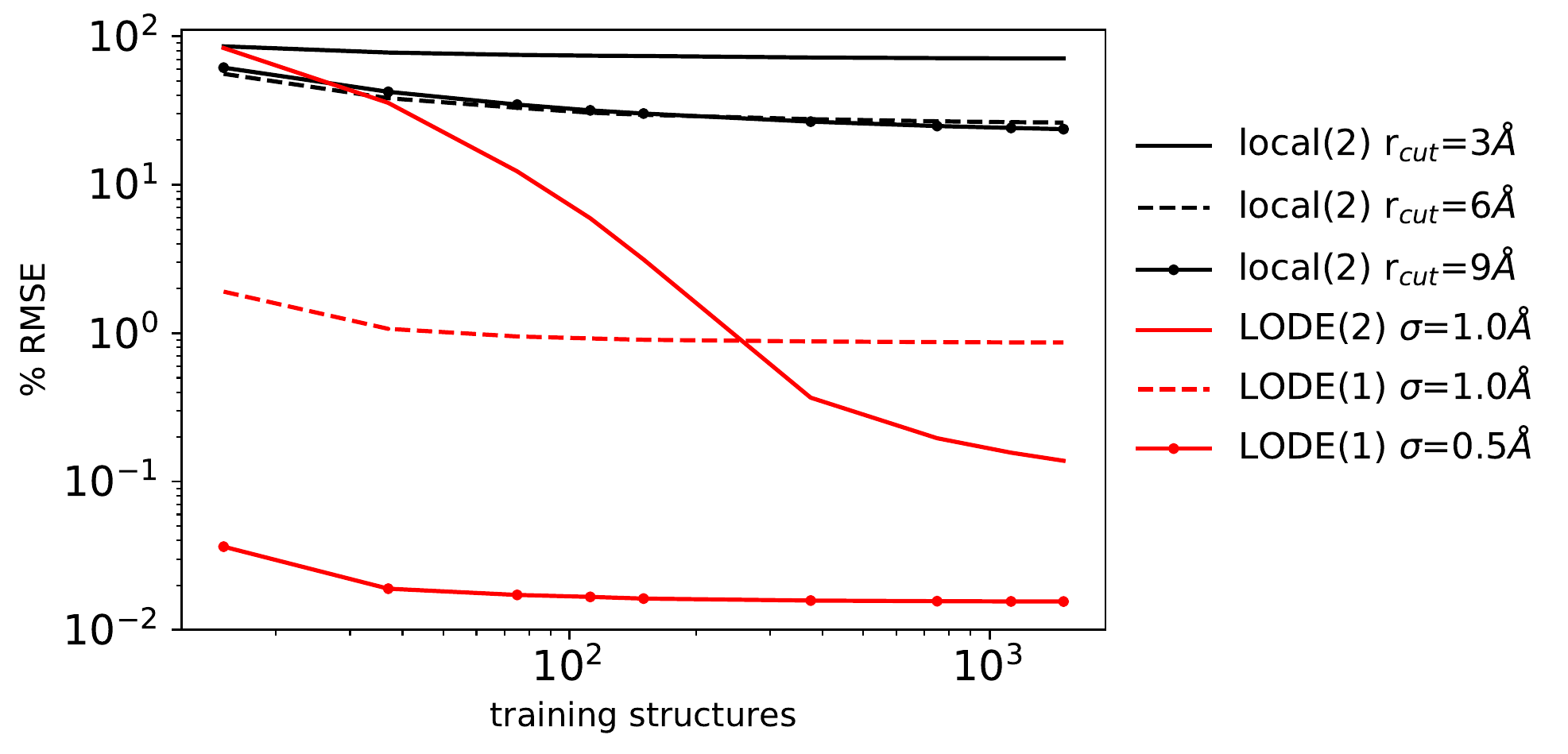}
    \caption{Learning curves for the electrostatic energy of an idealized random gas of point charges. The model is trained on 1500 randomly selected configurations and tested on other 500 independent configurations. (\textit{black full and dashed lines}) Local ML (SOAP) results at environment cutoffs of 3, 6 and 9~{\AA}. (\textit{red lines}) LODE($\nu=1$)  results at an environment cutoff of 2~{\AA} and Gaussian smearing of 0.5 and 1.0~{\AA}, and LODE($\nu=2$) results with a cutoff of 3{\AA}.\label{fig:random_nacl}}
\end{figure}

\subsection{A gas of point charges}

We begin by considering a toy system made of randomly distributed point-charges in a cubic box that is infinitely repeated in the three dimensions using periodic boundary conditions. The number of positive charges is equal to the number of negative charges, so that the system is overall neutral.
To limit the amplitude of energy fluctuations, we discard configurations in which two charges are closer together than~2.5~{\AA}. 
Following these prescriptions, we generate a total of 2000 configurations, each of which contains 64 atoms in cubic boxes spanning a broad range of densities, with side lengths between 12 and 20~{\AA}. 
For each of these configurations, we compute the electrostatic energy using the Ewald method, as implemented in LAMMPS~\cite{Plimpton1995}. Fig.~\ref{fig:random_nacl} compares the learning performance obtained using a local SOAP representation with different cutoffs, to the one obtained by direct application of the LODE representation.  In both cases, a Gaussian width of $\sigma$=1.0~{\AA} has been used to construct the density distribution of Eq.~\eqref{eq:rA-ket}.

The figure clearly demonstrate the inefficiency of a local model when attempting to learn a property that is dominated by long-range effects. Given that the training set contains few configurations with atoms closer than 3~{\AA}, the model with $r_\text{cut}$=3~{\AA} is almost completely ineffective. Even increasing the cutoff up to 9~{\AA}, a SOAP model barely reaches an accuracy of about 20\% RMSE when using the maximum number of training structures.
A linear model built using the LODE($\nu=1$) representation, on the other hand yields an  error below 1\%{} by using a handful of training points. 
As discussed above, this model represents exactly Coulomb interactions between fixed point charges, and the only reason the error does not converge to zero is the fact we use a Gaussian smearing in the definition of LODE, rather than $\delta$ distributions. This is apparent in the dramatic reduction of the error when halving the value of $\sigma$. 
A LODE($\nu=2$) model, although initially less effective, possesses sufficient descriptive power to reach, and then overcome, the accuracy of the linear $\nu=1,\sigma=1${\AA} model.
This simple example highlights how difficult it is to incorporate long-range physics with a conventional local structure representation, and demonstrates that the LODE features can, on their own, be used as a very efficient description to predict the electrostatic energy of a system of fixed point charges. 

\begin{figure*}[bhtb]
    \centering
    \includegraphics[width=0.9\linewidth]{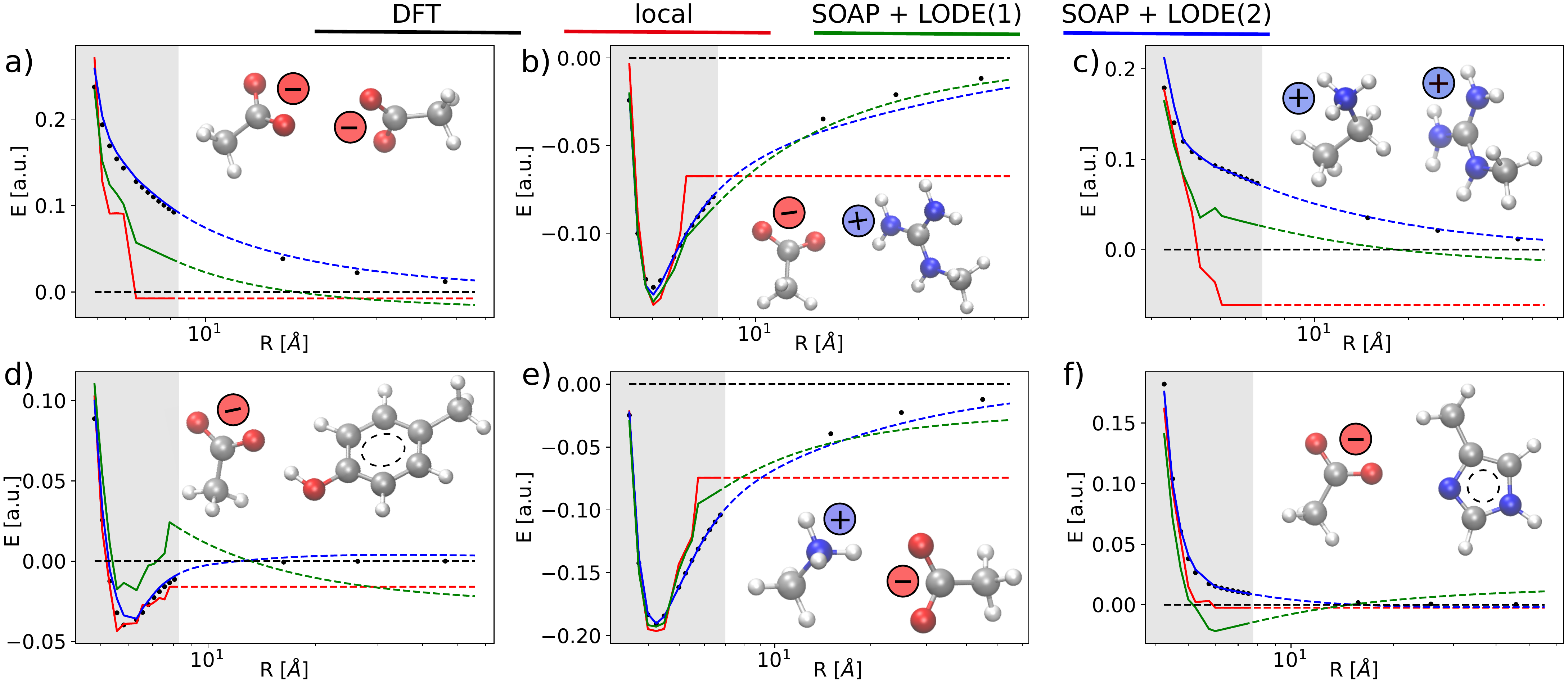}
     \caption{Comparison of reference and predicted binding curves of six molecular dimers.  (\textit{black dots}) DFT reference calculations, (\textit{red lines}) local SOAP predictions, (\textit{green lines}) combined SOAP and LODE(1) predictions, (\textit{blue lines}) combined SOAP and LODE(2) predictions. Full lines and shaded background represent the range of distances that is comparable to the geometries included in the training set. Dashed lines refer to predictions carried out in an extrapolative (long-range) regime. \label{fig:binding_curves} } 
\end{figure*}

\subsection{Binding curves of charged dimers}

We now consider a more realistic scenario, namely the problem of predicting the binding curves of a dataset of organic molecular dimers that carry an electric charge. We extract 661 different dimers containing H, C, N and O atoms  from the BioFragment Database (BFDb)~\cite{Burns2017}, where at least one of the two monomers in each dimer configuration has a net charge. 
This choice ensures that we focus the exercise on a problem for which permanent electrostatic interactions play a prominent role. Contrary to the NaCl toy system, however, one cannot expect that a fixed point-charge model would suffice to predict the binding curves. 
The dataset contains a multitude of chemical moieties, including neutral polar fragments, highly polarizable groups, and provides a realistic assessment of how well a LODE model can perform in practice.
For each of the 661 dimers, we consider 13 configurations where the reciprocal distance between the two monomers, defined as the distance between their geometric centers, spans an interval that can go from a minimum of $\sim$3~{\AA} to a maximum of $\sim$8~{\AA}. 
For each of these configurations, unrelaxed binding curves are computed at the DFT/B3LYP level using the FHI-aims quantum-chemistry package~\cite{Blum2009}. 
The training dataset is defined by considering the binding curves of the first 600 dimers out of the total of 661, while predictions are tested on the remaining 61. We also include the isolated monomers in the training set, so that the ML model has knowledge of the dissociation limit, and compute a few additional reference energies at larger separations, which are however not used for training. 
SOAP and LODE representations are defined within spherical environments of $r_\text{cut}=3.0$~{\AA}, while the Gaussian width of the density field is chosen to be $\sigma$=0.3 and 1.0~{\AA} respectively.

Before carrying out the learning exercise, the reference DFT energies are baselined with respect to the monomer energies, so that the model only has to reproduce the interaction energies between the two fragments. 
Upon this baselining, we find that optimal SOAP performances correspond to a RMSE $\sim$20\%, whereas a suitable combination between SOAP and LODE($\nu=2$) allows us to bring the error down to $\sim$4\%.
This substantial improvement can be justified by the large discrepancy between the SOAP and SOAP+LODE accuracy in representing the interaction between the monomers at intermediate and large distance. 
To clarify the issue further, we plot in Fig.~\ref{fig:binding_curves}  the predicted binding curves of 6 test dimers, against the reference DFT calculations. 
We observe that a SOAP-based local description is overall able to capture the short-range interactions with good accuracy. However, it becomes less and less effective as the distance between the monomers increases, to the point of being completely blind to changes in interatomic distances when the environments cutoff distance is overcome. Note that the performance of the local model at small separations is degraded substantially by the inclusion of fully dissociated dimers in the training set, because the representation cannot distinguish these configurations from those barely beyond the cutoff distance, that correspond to a non-zero value of the binding curve. 
The SOAP+LODE multiscale description, in contrast, can recognize the changes in separation between the monomers, leading to a smooth asymptotic behavior of the predicted binding curve. 
Although a linear model incorporating LODE($\nu=1$) allows us to halve the error made by SOAP down to $\sim$10\%, it is not sufficiently expressive to achieve predictive accuracy - particularly for binding curves that involve neutral monomers that do not have a $1/r$ asymptotic behavior. 

This limitation can be addressed using a non-linear kernel based on SOAP+LODE($\nu=2$). The resulting model is able to accurately predict the binding curves in the entire domain of distances, demonstrating its transferability across a vast spectrum of different chemical species and intermolecular configurations.
This is particularly remarkable, as the SOAP+LODE($\nu=2$) model does not only predict accurately systems that are dominated by monopole electrostatics (Fig.3-(a,b,c,e)), but also systems in which only one of the molecules is charged, and so interactions involve polarization as well as charge-dipole electrostatics  (Fig.3-(d,f)).
It should be noted, however, that the current scheme cannot transparently describe the physics of polarization or charge transfer. While the use of a composite SOAP+LODE kernel can describe how the environment of an atom affects its response to an external field, there is no explicit provision to represent how the field generated by far-away atoms depends on their neighboring structure.

\subsection{Dielectric response of liquid water}

\begin{figure}[tb]
    \centering
    \includegraphics[width=9cm]{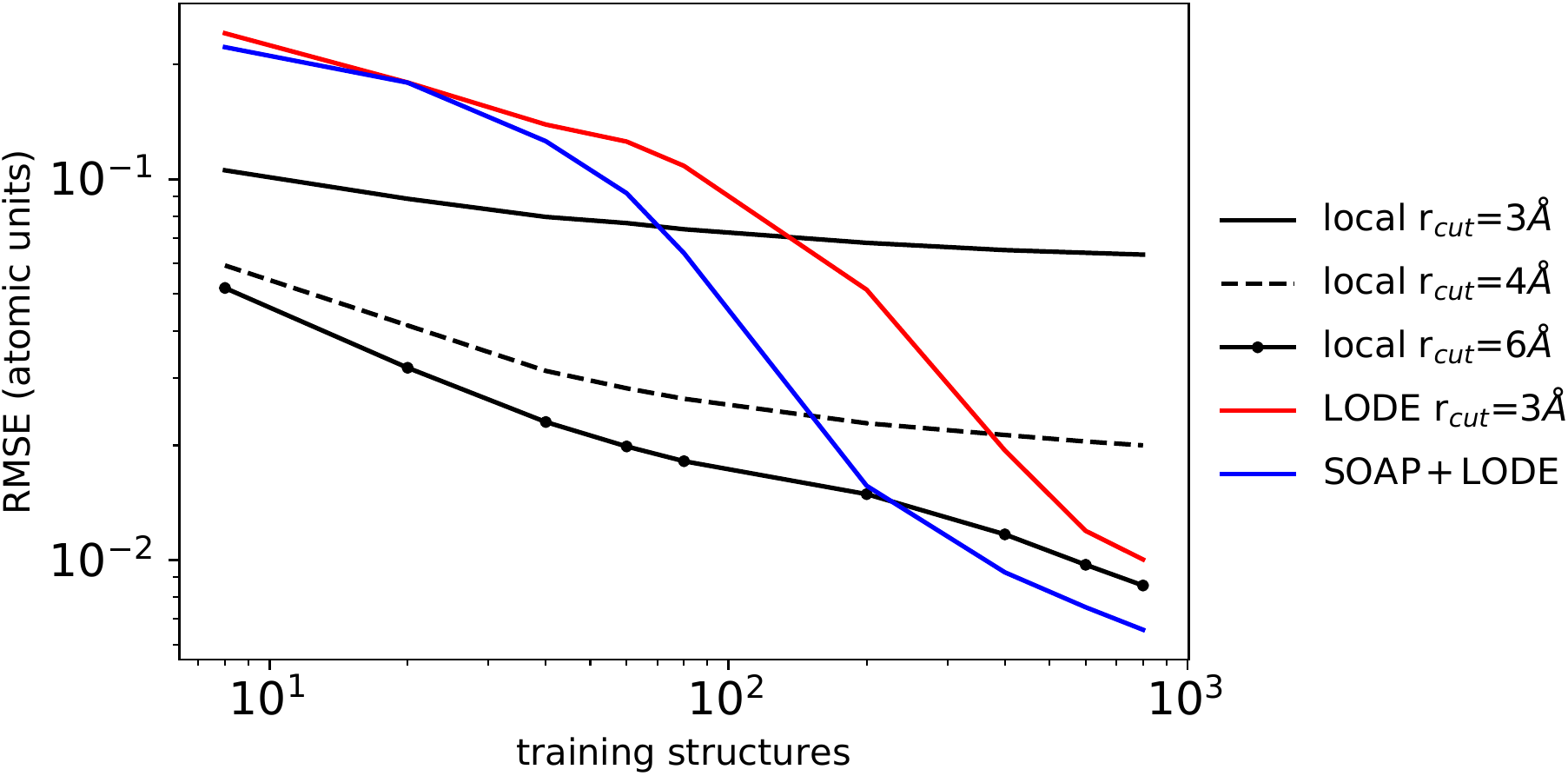}
   \caption{Learning curves for the isotropic component of the dielectric response tensor $\boldsymbol{\varepsilon}_\infty$ of liquid water. The model is trained on up to 800 randomly selected configurations and tested on other 200 independent configurations. (\textit{black full and dashed lines}) SOAP results with $r_\text{cut}$= 3, 4 and 6~{\AA}. (\textit{red line}) LODE results with $r_\text{cut}$=3~{\AA}. (\textit{blue line}) combined results of SOAP and LODE, both using $r_\text{cut}$=3~{\AA}.\label{fig:eps0_water}} 
\end{figure}

As a final example, we revisit the problem of constructing a model of the infinite-frequency dielectric response tensor $\boldsymbol{\varepsilon}_\infty$ of liquid water. Details about the dataset generation and the computation of the dielectric tensors are reported in Ref.~\cite{grisafi2018}.
In that work, we argued that a local model was inefficient in learning dielectric response because of its collective nature, and showed that using the Clausius-Mossotti relationship to map $\boldsymbol{\varepsilon}_\infty$ to more local quantities was greatly improving the model.
Here, LODE learning performances are only tested for the isotropic component of the tensor $\varepsilon_0=\text{Tr}[\boldsymbol{\varepsilon}_\infty]$, which was shown to be most sensitive to the collective nature of the physics of dielectrics. 
Similarly to the case of th BFDb, we use a non-linear kernel that combines a SOAP representations computed using an optimal Gaussian width of $\sigma$=0.3~{\AA}, and LODE($\nu=2$) features constructed starting from a Gaussian density of  $\sigma$=1.0~{\AA}.
Figure~\ref{fig:eps0_water} reports results obtained when learning on 800 randomly selected structures and predicting on other 200 independent configurations. 

Similarly to what has been observed in the previous example, LODE performs much better than SOAP when relying upon a local description of $r_\text{cut}$=3~{\AA}. In this case, however, we observe a substantial improvement of the performance of  SOAP when increasing the size of the local environments, eventually overcoming the LODE accuracy with a radial cutoff of $r_\text{cut}$=6~{\AA}. 
This might be a consequence of a less pronounced contribution of long-range tails, or - likely - of the fact that a cutoff of 6~{\AA} encompasses the entirety of the supercell, and therefore effectively provides a complete description of the input space of this specific dataset. 
Optimal ML predictions can be obtained when combining the fine-grained local description of SOAP at  $r_\text{cut}$=3~{\AA} with the coarse-grained and non-local description of LODE at the same cutoff. 
This behaviour highlights the multiscale character of $\varepsilon_0$, meaning that both the local many-body information and the long-range electrostatic effects need to be considered to get accurate predictions. 
It is also important to stress that a combination of SOAP and LODE is not only beneficial in terms of learning performance, but can also reduce the computational effort in evaluating the feature vector -- much like efficient methods for evaluating empirical potentials often treat separately short-range and long-range interactions. %

\section*{Conclusions}

Machine-learning of atomic-scale properties that are dominated by short-range interactions has reached a stage of maturity, with a substantial consensus about the ingredients of a successful model. 
The most commonly used frameworks incorporate symmetries and physical principles into the representation of atomic configurations, and achieve transferability by building additive property models.  
Furthermore, there is a growing understanding of the deep connections that exist between many of these methods, which is reflected in the fact that in most applications they reach similar levels of accuracy. 
In this paper we show how to extend these schemes in a way that makes it possible to incorporate long-range physics, without sacrificing the transferability of additive property models and the general applicability of rather abstract measures of atomic structure correlations.
The crux lies in the definition of an atom-density potential that folds global information on the structure and composition of a system into a local representation, that (1) has a physically-motivated asymptotic behavior with inter-atomic separation and (2) can be efficiently computed in a symmetry-consistent fashion using similar ideas as those that underlie the SOAP framework and related approaches. 

We apply this long-distance equivariant (LODE) representation focusing on the version that is based on a Coulomb-like atom-density potential. We demonstrate that, alone or in combination with SOAP, it outperforms local machine-learning methods in capturing long-range physics, for tasks that involve learning the electrostatic energy of a point-charge model, the binding curve of dimers of electrically charged organic fragments, and the dielectric constant of bulk water. 

These examples are little more than an assay that proves that this scheme can incorporate efficiently long-range information in atomistic machine learning. 
More work is needed to draw a systematic, formal connection between a ML model built on LODE features and long-range interatomic potentials, much like a connection has been shown between linear models built on density-based features and short-range many-body potentials~\cite{Glielmo2018,willatt2019,Drautz2019}; whether choosing other exponents in $\bra{\br}\ket{\CVp}$ can improve models of dispersion and of long-range effects that do not imply a characteristic asymptotic behavior; whether equivariant local features can be obtained by combining the expansion of the density and that of $\bra{\br}\ket{\CVp}$; whether the combination of SOAP and LODE can be used to improve the accuracy and the computational efficiency of existing ML forcefields; whether it is possible to incorporate polarizable atoms physics into the LODE framework.
Future investigation will address these and many other questions, and unearth the full potential of this physics-inspired approach to atomistic machine learning.

\section*{Acknowledgments}

The Authors would like to thank Clemence Corminboeuf and G\'abor Cs\'any for insightful comments on an early version of the manuscript. 
M.C and A.G. were supported by the European Research Council under the European Union's Horizon 2020 research and innovation programme (grant agreement no. 677013-HBMAP), and by the NCCR MARVEL, funded by the Swiss National Science Foundation. A.G. acknowledges funding by the MPG-EPFL Center for Molecular Nanoscience and Technology. 
We thank CSCS for providing CPU time under project id s843.


\begin{thebibliography}{47}%
\makeatletter
\providecommand \@ifxundefined [1]{%
 \@ifx{#1\undefined}
}%
\providecommand \@ifnum [1]{%
 \ifnum #1\expandafter \@firstoftwo
 \else \expandafter \@secondoftwo
 \fi
}%
\providecommand \@ifx [1]{%
 \ifx #1\expandafter \@firstoftwo
 \else \expandafter \@secondoftwo
 \fi
}%
\providecommand \natexlab [1]{#1}%
\providecommand \enquote  [1]{``#1''}%
\providecommand \bibnamefont  [1]{#1}%
\providecommand \bibfnamefont [1]{#1}%
\providecommand \citenamefont [1]{#1}%
\providecommand \href@noop [0]{\@secondoftwo}%
\providecommand \href [0]{\begingroup \@sanitize@url \@href}%
\providecommand \@href[1]{\@@startlink{#1}\@@href}%
\providecommand \@@href[1]{\endgroup#1\@@endlink}%
\providecommand \@sanitize@url [0]{\catcode `\\12\catcode `\$12\catcode
  `\&12\catcode `\#12\catcode `\^12\catcode `\_12\catcode `\%12\relax}%
\providecommand \@@startlink[1]{}%
\providecommand \@@endlink[0]{}%
\providecommand \url  [0]{\begingroup\@sanitize@url \@url }%
\providecommand \@url [1]{\endgroup\@href {#1}{\urlprefix }}%
\providecommand \urlprefix  [0]{URL }%
\providecommand \Eprint [0]{\href }%
\providecommand \doibase [0]{http://dx.doi.org/}%
\providecommand \selectlanguage [0]{\@gobble}%
\providecommand \bibinfo  [0]{\@secondoftwo}%
\providecommand \bibfield  [0]{\@secondoftwo}%
\providecommand \translation [1]{[#1]}%
\providecommand \BibitemOpen [0]{}%
\providecommand \bibitemStop [0]{}%
\providecommand \bibitemNoStop [0]{.\EOS\space}%
\providecommand \EOS [0]{\spacefactor3000\relax}%
\providecommand \BibitemShut  [1]{\csname bibitem#1\endcsname}%
\let\auto@bib@innerbib\@empty
\bibitem [{\citenamefont {Behler}(2017)}]{behler2017}%
  \BibitemOpen
  \bibfield  {author} {\bibinfo {author} {\bibfnamefont {J.}~\bibnamefont
  {Behler}},\ }\href@noop {} {\bibfield  {journal} {\bibinfo  {journal}
  {Angewandte Chemie International Edition}\ }\textbf {\bibinfo {volume}
  {56}},\ \bibinfo {pages} {12828} (\bibinfo {year} {2017})}\BibitemShut
  {NoStop}%
\bibitem [{\citenamefont {Behler}\ and\ \citenamefont
  {Parrinello}(2007)}]{behler2007}%
  \BibitemOpen
  \bibfield  {author} {\bibinfo {author} {\bibfnamefont {J.}~\bibnamefont
  {Behler}}\ and\ \bibinfo {author} {\bibfnamefont {M.}~\bibnamefont
  {Parrinello}},\ }\href {\doibase 10.1103/PhysRevLett.98.146401} {\bibfield
  {journal} {\bibinfo  {journal} {Phys. Rev. Lett.}\ }\textbf {\bibinfo
  {volume} {98}},\ \bibinfo {pages} {146401} (\bibinfo {year}
  {2007})}\BibitemShut {NoStop}%
\bibitem [{\citenamefont {Bart\'ok}\ \emph {et~al.}(2013)\citenamefont
  {Bart\'ok}, \citenamefont {Kondor},\ and\ \citenamefont
  {Cs\'anyi}}]{bartok2013}%
  \BibitemOpen
  \bibfield  {author} {\bibinfo {author} {\bibfnamefont {A.~P.}\ \bibnamefont
  {Bart\'ok}}, \bibinfo {author} {\bibfnamefont {R.}~\bibnamefont {Kondor}}, \
  and\ \bibinfo {author} {\bibfnamefont {G.}~\bibnamefont {Cs\'anyi}},\
  }\href@noop {} {\bibfield  {journal} {\bibinfo  {journal} {Phys. Rev. B}\
  }\textbf {\bibinfo {volume} {87}},\ \bibinfo {pages} {184115} (\bibinfo
  {year} {2013})}\BibitemShut {NoStop}%
\bibitem [{\citenamefont {Shapeev}(2016)}]{Shapeev2015}%
  \BibitemOpen
  \bibfield  {author} {\bibinfo {author} {\bibfnamefont {A.}~\bibnamefont
  {Shapeev}},\ }\href@noop {} {\bibfield  {journal} {\bibinfo  {journal}
  {Multiscale Model. Sim.}\ }\textbf {\bibinfo {volume} {14}},\ \bibinfo
  {pages} {1153} (\bibinfo {year} {2016})}\BibitemShut {NoStop}%
\bibitem [{\citenamefont {Glielmo}\ \emph {et~al.}(2017)\citenamefont
  {Glielmo}, \citenamefont {Sollich},\ and\ \citenamefont
  {De~Vita}}]{Glielmo2017}%
  \BibitemOpen
  \bibfield  {author} {\bibinfo {author} {\bibfnamefont {A.}~\bibnamefont
  {Glielmo}}, \bibinfo {author} {\bibfnamefont {P.}~\bibnamefont {Sollich}}, \
  and\ \bibinfo {author} {\bibfnamefont {A.}~\bibnamefont {De~Vita}},\
  }\href@noop {} {\bibfield  {journal} {\bibinfo  {journal} {Phys. Rev. B}\
  }\textbf {\bibinfo {volume} {95}},\ \bibinfo {pages} {214302} (\bibinfo
  {year} {2017})}\BibitemShut {NoStop}%
\bibitem [{\citenamefont {Grisafi}\ \emph {et~al.}(2018)\citenamefont
  {Grisafi}, \citenamefont {Wilkins}, \citenamefont {Cs\'anyi},\ and\
  \citenamefont {Ceriotti}}]{grisafi2018}%
  \BibitemOpen
  \bibfield  {author} {\bibinfo {author} {\bibfnamefont {A.}~\bibnamefont
  {Grisafi}}, \bibinfo {author} {\bibfnamefont {D.~M.}\ \bibnamefont
  {Wilkins}}, \bibinfo {author} {\bibfnamefont {G.}~\bibnamefont {Cs\'anyi}}, \
  and\ \bibinfo {author} {\bibfnamefont {M.}~\bibnamefont {Ceriotti}},\
  }\href@noop {} {\bibfield  {journal} {\bibinfo  {journal} {Phys. Rev. Lett.}\
  }\textbf {\bibinfo {volume} {120}},\ \bibinfo {pages} {036002} (\bibinfo
  {year} {2018})}\BibitemShut {NoStop}%
\bibitem [{\citenamefont {Bart{\'o}k}\ \emph {et~al.}(2017)\citenamefont
  {Bart{\'o}k}, \citenamefont {De}, \citenamefont {Poelking}, \citenamefont
  {Bernstein}, \citenamefont {Kermode}, \citenamefont {Cs{\'a}nyi},\ and\
  \citenamefont {Ceriotti}}]{bartok2017}%
  \BibitemOpen
  \bibfield  {author} {\bibinfo {author} {\bibfnamefont {A.~P.}\ \bibnamefont
  {Bart{\'o}k}}, \bibinfo {author} {\bibfnamefont {S.}~\bibnamefont {De}},
  \bibinfo {author} {\bibfnamefont {C.}~\bibnamefont {Poelking}}, \bibinfo
  {author} {\bibfnamefont {N.}~\bibnamefont {Bernstein}}, \bibinfo {author}
  {\bibfnamefont {J.~R.}\ \bibnamefont {Kermode}}, \bibinfo {author}
  {\bibfnamefont {G.}~\bibnamefont {Cs{\'a}nyi}}, \ and\ \bibinfo {author}
  {\bibfnamefont {M.}~\bibnamefont {Ceriotti}},\ }\href@noop {} {\bibfield
  {journal} {\bibinfo  {journal} {Sci. Adv.}\ }\textbf {\bibinfo {volume} {3}}
  (\bibinfo {year} {2017})}\BibitemShut {NoStop}%
\bibitem [{\citenamefont {Chmiela}\ \emph {et~al.}(2017)\citenamefont
  {Chmiela}, \citenamefont {Tkatchenko}, \citenamefont {Sauceda}, \citenamefont
  {Poltavsky}, \citenamefont {Sch{\"u}tt},\ and\ \citenamefont
  {M{\"u}ller}}]{Chmiela2017}%
  \BibitemOpen
  \bibfield  {author} {\bibinfo {author} {\bibfnamefont {S.}~\bibnamefont
  {Chmiela}}, \bibinfo {author} {\bibfnamefont {A.}~\bibnamefont {Tkatchenko}},
  \bibinfo {author} {\bibfnamefont {H.~E.}\ \bibnamefont {Sauceda}}, \bibinfo
  {author} {\bibfnamefont {I.}~\bibnamefont {Poltavsky}}, \bibinfo {author}
  {\bibfnamefont {K.~T.}\ \bibnamefont {Sch{\"u}tt}}, \ and\ \bibinfo {author}
  {\bibfnamefont {K.-R.}\ \bibnamefont {M{\"u}ller}},\ }\href@noop {}
  {\bibfield  {journal} {\bibinfo  {journal} {Sci. Adv.}\ }\textbf {\bibinfo
  {volume} {3}} (\bibinfo {year} {2017})}\BibitemShut {NoStop}%
\bibitem [{\citenamefont {Zhang}\ \emph {et~al.}(2018)\citenamefont {Zhang},
  \citenamefont {Han}, \citenamefont {Wang}, \citenamefont {Car},\ and\
  \citenamefont {E}}]{Zhang2018}%
  \BibitemOpen
  \bibfield  {author} {\bibinfo {author} {\bibfnamefont {L.}~\bibnamefont
  {Zhang}}, \bibinfo {author} {\bibfnamefont {J.}~\bibnamefont {Han}}, \bibinfo
  {author} {\bibfnamefont {H.}~\bibnamefont {Wang}}, \bibinfo {author}
  {\bibfnamefont {R.}~\bibnamefont {Car}}, \ and\ \bibinfo {author}
  {\bibfnamefont {W.}~\bibnamefont {E}},\ }\href@noop {} {\bibfield  {journal}
  {\bibinfo  {journal} {Phys. Rev. Lett.}\ }\textbf {\bibinfo {volume} {120}},\
  \bibinfo {pages} {143001} (\bibinfo {year} {2018})}\BibitemShut {NoStop}%
\bibitem [{\citenamefont {Prodan}\ and\ \citenamefont
  {Kohn}(2005)}]{Prodan2005}%
  \BibitemOpen
  \bibfield  {author} {\bibinfo {author} {\bibfnamefont {E.}~\bibnamefont
  {Prodan}}\ and\ \bibinfo {author} {\bibfnamefont {W.}~\bibnamefont {Kohn}},\
  }\href@noop {} {\bibfield  {journal} {\bibinfo  {journal} {Proceedings of the
  National Academy of Sciences}\ }\textbf {\bibinfo {volume} {102}},\ \bibinfo
  {pages} {11635} (\bibinfo {year} {2005})}\BibitemShut {NoStop}%
\bibitem [{\citenamefont {Kjellander}(2018)}]{kjellander2018}%
  \BibitemOpen
  \bibfield  {author} {\bibinfo {author} {\bibfnamefont {R.}~\bibnamefont
  {Kjellander}},\ }\href@noop {} {\bibfield  {journal} {\bibinfo  {journal}
  {The Journal of Chemical Physics}\ }\textbf {\bibinfo {volume} {148}},\
  \bibinfo {pages} {193701} (\bibinfo {year} {2018})}\BibitemShut {NoStop}%
\bibitem [{\citenamefont {Guo}\ \emph {et~al.}(2018)\citenamefont {Guo},
  \citenamefont {Ambrosio}, \citenamefont {Chen}, \citenamefont {Gono},\ and\
  \citenamefont {Pasquarello}}]{guo2018}%
  \BibitemOpen
  \bibfield  {author} {\bibinfo {author} {\bibfnamefont {Z.}~\bibnamefont
  {Guo}}, \bibinfo {author} {\bibfnamefont {F.}~\bibnamefont {Ambrosio}},
  \bibinfo {author} {\bibfnamefont {W.}~\bibnamefont {Chen}}, \bibinfo {author}
  {\bibfnamefont {P.}~\bibnamefont {Gono}}, \ and\ \bibinfo {author}
  {\bibfnamefont {A.}~\bibnamefont {Pasquarello}},\ }\href@noop {} {\bibfield
  {journal} {\bibinfo  {journal} {Chemistry of Materials}\ }\textbf {\bibinfo
  {volume} {30}},\ \bibinfo {pages} {94} (\bibinfo {year} {2018})}\BibitemShut
  {NoStop}%
\bibitem [{\citenamefont {Jorn}\ \emph {et~al.}(2013)\citenamefont {Jorn},
  \citenamefont {Kumar}, \citenamefont {Abraham},\ and\ \citenamefont
  {Voth}}]{jorn2013}%
  \BibitemOpen
  \bibfield  {author} {\bibinfo {author} {\bibfnamefont {R.}~\bibnamefont
  {Jorn}}, \bibinfo {author} {\bibfnamefont {R.}~\bibnamefont {Kumar}},
  \bibinfo {author} {\bibfnamefont {D.~P.}\ \bibnamefont {Abraham}}, \ and\
  \bibinfo {author} {\bibfnamefont {G.~A.}\ \bibnamefont {Voth}},\ }\href@noop
  {} {\bibfield  {journal} {\bibinfo  {journal} {The Journal of Physical
  Chemistry C}\ }\textbf {\bibinfo {volume} {117}},\ \bibinfo {pages} {3747}
  (\bibinfo {year} {2013})}\BibitemShut {NoStop}%
\bibitem [{\citenamefont {French}\ \emph {et~al.}(2010)\citenamefont {French},
  \citenamefont {Parsegian}, \citenamefont {Podgornik}, \citenamefont {Rajter},
  \citenamefont {Jagota}, \citenamefont {Luo}, \citenamefont {Asthagiri},
  \citenamefont {Chaudhury}, \citenamefont {Chiang}, \citenamefont {Granick},
  \citenamefont {Kalinin}, \citenamefont {Kardar}, \citenamefont {Kjellander},
  \citenamefont {Langreth}, \citenamefont {Lewis}, \citenamefont {Lustig},
  \citenamefont {Wesolowski}, \citenamefont {Wettlaufer}, \citenamefont
  {Ching}, \citenamefont {Finnis}, \citenamefont {Houlihan}, \citenamefont {von
  Lilienfeld}, \citenamefont {van Oss},\ and\ \citenamefont
  {Zemb}}]{french2010}%
  \BibitemOpen
  \bibfield  {author} {\bibinfo {author} {\bibfnamefont {R.~H.}\ \bibnamefont
  {French}}, \bibinfo {author} {\bibfnamefont {V.~A.}\ \bibnamefont
  {Parsegian}}, \bibinfo {author} {\bibfnamefont {R.}~\bibnamefont
  {Podgornik}}, \bibinfo {author} {\bibfnamefont {R.~F.}\ \bibnamefont
  {Rajter}}, \bibinfo {author} {\bibfnamefont {A.}~\bibnamefont {Jagota}},
  \bibinfo {author} {\bibfnamefont {J.}~\bibnamefont {Luo}}, \bibinfo {author}
  {\bibfnamefont {D.}~\bibnamefont {Asthagiri}}, \bibinfo {author}
  {\bibfnamefont {M.~K.}\ \bibnamefont {Chaudhury}}, \bibinfo {author}
  {\bibfnamefont {Y.-m.}\ \bibnamefont {Chiang}}, \bibinfo {author}
  {\bibfnamefont {S.}~\bibnamefont {Granick}}, \bibinfo {author} {\bibfnamefont
  {S.}~\bibnamefont {Kalinin}}, \bibinfo {author} {\bibfnamefont
  {M.}~\bibnamefont {Kardar}}, \bibinfo {author} {\bibfnamefont
  {R.}~\bibnamefont {Kjellander}}, \bibinfo {author} {\bibfnamefont {D.~C.}\
  \bibnamefont {Langreth}}, \bibinfo {author} {\bibfnamefont {J.}~\bibnamefont
  {Lewis}}, \bibinfo {author} {\bibfnamefont {S.}~\bibnamefont {Lustig}},
  \bibinfo {author} {\bibfnamefont {D.}~\bibnamefont {Wesolowski}}, \bibinfo
  {author} {\bibfnamefont {J.~S.}\ \bibnamefont {Wettlaufer}}, \bibinfo
  {author} {\bibfnamefont {W.-Y.}\ \bibnamefont {Ching}}, \bibinfo {author}
  {\bibfnamefont {M.}~\bibnamefont {Finnis}}, \bibinfo {author} {\bibfnamefont
  {F.}~\bibnamefont {Houlihan}}, \bibinfo {author} {\bibfnamefont {O.~A.}\
  \bibnamefont {von Lilienfeld}}, \bibinfo {author} {\bibfnamefont {C.~J.}\
  \bibnamefont {van Oss}}, \ and\ \bibinfo {author} {\bibfnamefont
  {T.}~\bibnamefont {Zemb}},\ }\href@noop {} {\bibfield  {journal} {\bibinfo
  {journal} {Rev. Mod. Phys.}\ }\textbf {\bibinfo {volume} {82}},\ \bibinfo
  {pages} {1887} (\bibinfo {year} {2010})}\BibitemShut {NoStop}%
\bibitem [{\citenamefont {Bart\'ok}\ \emph {et~al.}(2010)\citenamefont
  {Bart\'ok}, \citenamefont {Payne}, \citenamefont {Kondor},\ and\
  \citenamefont {Cs\'anyi}}]{bartok2010}%
  \BibitemOpen
  \bibfield  {author} {\bibinfo {author} {\bibfnamefont {A.~P.}\ \bibnamefont
  {Bart\'ok}}, \bibinfo {author} {\bibfnamefont {M.~C.}\ \bibnamefont {Payne}},
  \bibinfo {author} {\bibfnamefont {R.}~\bibnamefont {Kondor}}, \ and\ \bibinfo
  {author} {\bibfnamefont {G.}~\bibnamefont {Cs\'anyi}},\ }\href@noop {}
  {\bibfield  {journal} {\bibinfo  {journal} {Phys. Rev. Lett.}\ }\textbf
  {\bibinfo {volume} {104}},\ \bibinfo {pages} {136403} (\bibinfo {year}
  {2010})}\BibitemShut {NoStop}%
\bibitem [{\citenamefont {Deng}\ \emph {et~al.}(2019)\citenamefont {Deng},
  \citenamefont {Chen}, \citenamefont {Li},\ and\ \citenamefont
  {Ong}}]{Deng2019}%
  \BibitemOpen
  \bibfield  {author} {\bibinfo {author} {\bibfnamefont {Z.}~\bibnamefont
  {Deng}}, \bibinfo {author} {\bibfnamefont {C.}~\bibnamefont {Chen}}, \bibinfo
  {author} {\bibfnamefont {X.-G.}\ \bibnamefont {Li}}, \ and\ \bibinfo {author}
  {\bibfnamefont {S.~P.}\ \bibnamefont {Ong}},\ }\href@noop {} {\bibfield
  {journal} {\bibinfo  {journal} {npj Computational Materials}\ }\textbf
  {\bibinfo {volume} {5}},\ \bibinfo {pages} {75} (\bibinfo {year}
  {2019})}\BibitemShut {NoStop}%
\bibitem [{\citenamefont {Artrith}\ \emph {et~al.}(2011)\citenamefont
  {Artrith}, \citenamefont {Morawietz},\ and\ \citenamefont
  {Behler}}]{Artrith2011}%
  \BibitemOpen
  \bibfield  {author} {\bibinfo {author} {\bibfnamefont {N.}~\bibnamefont
  {Artrith}}, \bibinfo {author} {\bibfnamefont {T.}~\bibnamefont {Morawietz}},
  \ and\ \bibinfo {author} {\bibfnamefont {J.}~\bibnamefont {Behler}},\ }\href
  {\doibase 10.1103/PhysRevB.83.153101} {\bibfield  {journal} {\bibinfo
  {journal} {Phys. Rev. B}\ }\textbf {\bibinfo {volume} {83}},\ \bibinfo
  {pages} {153101} (\bibinfo {year} {2011})}\BibitemShut {NoStop}%
\bibitem [{\citenamefont {Bereau}\ \emph {et~al.}(2015)\citenamefont {Bereau},
  \citenamefont {Andrienko},\ and\ \citenamefont {von
  Lilienfeld}}]{Bereau2015}%
  \BibitemOpen
  \bibfield  {author} {\bibinfo {author} {\bibfnamefont {T.}~\bibnamefont
  {Bereau}}, \bibinfo {author} {\bibfnamefont {D.}~\bibnamefont {Andrienko}}, \
  and\ \bibinfo {author} {\bibfnamefont {O.~A.}\ \bibnamefont {von
  Lilienfeld}},\ }\href@noop {} {\bibfield  {journal} {\bibinfo  {journal} {J.
  Chem. Theory Comput.}\ }\textbf {\bibinfo {volume} {11}},\ \bibinfo {pages}
  {3225} (\bibinfo {year} {2015})}\BibitemShut {NoStop}%
\bibitem [{\citenamefont {Bereau}\ \emph {et~al.}(2018)\citenamefont {Bereau},
  \citenamefont {DiStasio}, \citenamefont {Tkatchenko},\ and\ \citenamefont
  {von Lilienfeld}}]{Bereau2017}%
  \BibitemOpen
  \bibfield  {author} {\bibinfo {author} {\bibfnamefont {T.}~\bibnamefont
  {Bereau}}, \bibinfo {author} {\bibfnamefont {R.~A.}\ \bibnamefont
  {DiStasio}}, \bibinfo {author} {\bibfnamefont {A.}~\bibnamefont
  {Tkatchenko}}, \ and\ \bibinfo {author} {\bibfnamefont {O.~A.}\ \bibnamefont
  {von Lilienfeld}},\ }\href@noop {} {\bibfield  {journal} {\bibinfo  {journal}
  {J. Chem. Phys.}\ }\textbf {\bibinfo {volume} {148}},\ \bibinfo {pages}
  {241706} (\bibinfo {year} {2018})}\BibitemShut {NoStop}%
\bibitem [{\citenamefont {Bleiziffer}\ \emph {et~al.}(2018)\citenamefont
  {Bleiziffer}, \citenamefont {Schaller},\ and\ \citenamefont
  {Riniker}}]{Bleiziffer2018}%
  \BibitemOpen
  \bibfield  {author} {\bibinfo {author} {\bibfnamefont {P.}~\bibnamefont
  {Bleiziffer}}, \bibinfo {author} {\bibfnamefont {K.}~\bibnamefont
  {Schaller}}, \ and\ \bibinfo {author} {\bibfnamefont {S.}~\bibnamefont
  {Riniker}},\ }\href@noop {} {\bibfield  {journal} {\bibinfo  {journal}
  {Journal of Chemical Information and Modeling}\ }\textbf {\bibinfo {volume}
  {58}},\ \bibinfo {pages} {579} (\bibinfo {year} {2018})}\BibitemShut
  {NoStop}%
\bibitem [{\citenamefont {Nebgen}\ \emph {et~al.}(2018)\citenamefont {Nebgen},
  \citenamefont {Lubbers}, \citenamefont {Smith}, \citenamefont {Sifain},
  \citenamefont {Lokhov}, \citenamefont {Isayev}, \citenamefont {Roitberg},
  \citenamefont {Barros},\ and\ \citenamefont {Tretiak}}]{Nebgen2018}%
  \BibitemOpen
  \bibfield  {author} {\bibinfo {author} {\bibfnamefont {B.}~\bibnamefont
  {Nebgen}}, \bibinfo {author} {\bibfnamefont {N.}~\bibnamefont {Lubbers}},
  \bibinfo {author} {\bibfnamefont {J.~S.}\ \bibnamefont {Smith}}, \bibinfo
  {author} {\bibfnamefont {A.~E.}\ \bibnamefont {Sifain}}, \bibinfo {author}
  {\bibfnamefont {A.}~\bibnamefont {Lokhov}}, \bibinfo {author} {\bibfnamefont
  {O.}~\bibnamefont {Isayev}}, \bibinfo {author} {\bibfnamefont {A.~E.}\
  \bibnamefont {Roitberg}}, \bibinfo {author} {\bibfnamefont {K.}~\bibnamefont
  {Barros}}, \ and\ \bibinfo {author} {\bibfnamefont {S.}~\bibnamefont
  {Tretiak}},\ }\href@noop {} {\bibfield  {journal} {\bibinfo  {journal} {J.
  Chem. Theory Comput.}\ }\textbf {\bibinfo {volume} {14}},\ \bibinfo {pages}
  {4687} (\bibinfo {year} {2018})}\BibitemShut {NoStop}%
\bibitem [{\citenamefont {Yao}\ \emph {et~al.}(2018)\citenamefont {Yao},
  \citenamefont {Herr}, \citenamefont {Toth}, \citenamefont {Mckintyre},\ and\
  \citenamefont {Parkhill}}]{Yao2018}%
  \BibitemOpen
  \bibfield  {author} {\bibinfo {author} {\bibfnamefont {K.}~\bibnamefont
  {Yao}}, \bibinfo {author} {\bibfnamefont {J.~E.}\ \bibnamefont {Herr}},
  \bibinfo {author} {\bibfnamefont {D.}~\bibnamefont {Toth}}, \bibinfo {author}
  {\bibfnamefont {R.}~\bibnamefont {Mckintyre}}, \ and\ \bibinfo {author}
  {\bibfnamefont {J.}~\bibnamefont {Parkhill}},\ }\href {\doibase
  10.1039/C7SC04934J} {\bibfield  {journal} {\bibinfo  {journal} {Chem. Sci.}\
  }\textbf {\bibinfo {volume} {9}},\ \bibinfo {pages} {2261} (\bibinfo {year}
  {2018})}\BibitemShut {NoStop}%
\bibitem [{\citenamefont {Ghasemi}\ \emph {et~al.}(2015)\citenamefont
  {Ghasemi}, \citenamefont {Hofstetter}, \citenamefont {Saha},\ and\
  \citenamefont {Goedecker}}]{Ghasemi2015}%
  \BibitemOpen
  \bibfield  {author} {\bibinfo {author} {\bibfnamefont {S.~A.}\ \bibnamefont
  {Ghasemi}}, \bibinfo {author} {\bibfnamefont {A.}~\bibnamefont {Hofstetter}},
  \bibinfo {author} {\bibfnamefont {S.}~\bibnamefont {Saha}}, \ and\ \bibinfo
  {author} {\bibfnamefont {S.}~\bibnamefont {Goedecker}},\ }\href {\doibase
  10.1103/PhysRevB.92.045131} {\bibfield  {journal} {\bibinfo  {journal} {Phys.
  Rev. B}\ }\textbf {\bibinfo {volume} {92}},\ \bibinfo {pages} {045131}
  (\bibinfo {year} {2015})}\BibitemShut {NoStop}%
\bibitem [{\citenamefont {Faraji}\ \emph {et~al.}(2017)\citenamefont {Faraji},
  \citenamefont {Ghasemi}, \citenamefont {Rostami}, \citenamefont
  {Rasoulkhani}, \citenamefont {Schaefer}, \citenamefont {Goedecker},\ and\
  \citenamefont {Amsler}}]{Faraji2017}%
  \BibitemOpen
  \bibfield  {author} {\bibinfo {author} {\bibfnamefont {S.}~\bibnamefont
  {Faraji}}, \bibinfo {author} {\bibfnamefont {S.~A.}\ \bibnamefont {Ghasemi}},
  \bibinfo {author} {\bibfnamefont {S.}~\bibnamefont {Rostami}}, \bibinfo
  {author} {\bibfnamefont {R.}~\bibnamefont {Rasoulkhani}}, \bibinfo {author}
  {\bibfnamefont {B.}~\bibnamefont {Schaefer}}, \bibinfo {author}
  {\bibfnamefont {S.}~\bibnamefont {Goedecker}}, \ and\ \bibinfo {author}
  {\bibfnamefont {M.}~\bibnamefont {Amsler}},\ }\href@noop {} {\bibfield
  {journal} {\bibinfo  {journal} {Phys. Rev. B}\ }\textbf {\bibinfo {volume}
  {95}},\ \bibinfo {pages} {104105} (\bibinfo {year} {2017})}\BibitemShut
  {NoStop}%
\bibitem [{\citenamefont {B{\"o}ttcher}\ \emph {et~al.}(1978)\citenamefont
  {B{\"o}ttcher}, \citenamefont {van Belle}, \citenamefont {Bordewijk},\ and\
  \citenamefont {Rip}}]{bottcher1978}%
  \BibitemOpen
  \bibfield  {author} {\bibinfo {author} {\bibfnamefont {C.}~\bibnamefont
  {B{\"o}ttcher}}, \bibinfo {author} {\bibfnamefont {O.}~\bibnamefont {van
  Belle}}, \bibinfo {author} {\bibfnamefont {P.}~\bibnamefont {Bordewijk}}, \
  and\ \bibinfo {author} {\bibfnamefont {A.}~\bibnamefont {Rip}},\ }\href@noop
  {} {\emph {\bibinfo {title} {Theory of electric polarization}}}\ (\bibinfo
  {publisher} {Elsevier Scientific Pub. Co.},\ \bibinfo {year}
  {1978})\BibitemShut {NoStop}%
\bibitem [{\citenamefont {Resta}(1994)}]{resta1994}%
  \BibitemOpen
  \bibfield  {author} {\bibinfo {author} {\bibfnamefont {R.}~\bibnamefont
  {Resta}},\ }\href@noop {} {\bibfield  {journal} {\bibinfo  {journal} {Rev.
  Mod. Phys.}\ }\textbf {\bibinfo {volume} {66}},\ \bibinfo {pages} {899}
  (\bibinfo {year} {1994})}\BibitemShut {NoStop}%
\bibitem [{\citenamefont {Resta}(2010)}]{Resta2010}%
  \BibitemOpen
  \bibfield  {author} {\bibinfo {author} {\bibfnamefont {R.}~\bibnamefont
  {Resta}},\ }\href {http://stacks.iop.org/0953-8984/22/i=12/a=123201}
  {\bibfield  {journal} {\bibinfo  {journal} {Journal of Physics: Condensed
  Matter}\ }\textbf {\bibinfo {volume} {22}},\ \bibinfo {pages} {123201}
  (\bibinfo {year} {2010})}\BibitemShut {NoStop}%
\bibitem [{\citenamefont {Zhang}\ \emph {et~al.}(2019)\citenamefont {Zhang},
  \citenamefont {Chen}, \citenamefont {Wu}, \citenamefont {Wang}, \citenamefont
  {E},\ and\ \citenamefont {Car}}]{Zhang2019arxiv}%
  \BibitemOpen
  \bibfield  {author} {\bibinfo {author} {\bibfnamefont {L.}~\bibnamefont
  {Zhang}}, \bibinfo {author} {\bibfnamefont {M.}~\bibnamefont {Chen}},
  \bibinfo {author} {\bibfnamefont {X.}~\bibnamefont {Wu}}, \bibinfo {author}
  {\bibfnamefont {H.}~\bibnamefont {Wang}}, \bibinfo {author} {\bibfnamefont
  {W.}~\bibnamefont {E}}, \ and\ \bibinfo {author} {\bibfnamefont
  {R.}~\bibnamefont {Car}},\ }\href@noop {} {\bibfield  {journal} {\bibinfo
  {journal} {arXiv:1906.11434}\ } (\bibinfo {year} {2019})}\BibitemShut
  {NoStop}%
\bibitem [{\citenamefont {Wilkins}\ \emph {et~al.}(2019)\citenamefont
  {Wilkins}, \citenamefont {Grisafi}, \citenamefont {Yang}, \citenamefont
  {Lao}, \citenamefont {DiStasio},\ and\ \citenamefont
  {Ceriotti}}]{Wilkins2019}%
  \BibitemOpen
  \bibfield  {author} {\bibinfo {author} {\bibfnamefont {D.~M.}\ \bibnamefont
  {Wilkins}}, \bibinfo {author} {\bibfnamefont {A.}~\bibnamefont {Grisafi}},
  \bibinfo {author} {\bibfnamefont {Y.}~\bibnamefont {Yang}}, \bibinfo {author}
  {\bibfnamefont {K.~U.}\ \bibnamefont {Lao}}, \bibinfo {author} {\bibfnamefont
  {R.~A.}\ \bibnamefont {DiStasio}}, \ and\ \bibinfo {author} {\bibfnamefont
  {M.}~\bibnamefont {Ceriotti}},\ }\href {\doibase 10.1073/pnas.1816132116}
  {\bibfield  {journal} {\bibinfo  {journal} {Proc. Natl. Acad. Sci.}\ }\textbf
  {\bibinfo {volume} {116}},\ \bibinfo {pages} {3401} (\bibinfo {year}
  {2019})}\BibitemShut {NoStop}%
\bibitem [{\citenamefont {Rupp}\ \emph {et~al.}(2012)\citenamefont {Rupp},
  \citenamefont {Tkatchenko}, \citenamefont {M\"uller},\ and\ \citenamefont
  {von Lilienfeld}}]{Rupp2012}%
  \BibitemOpen
  \bibfield  {author} {\bibinfo {author} {\bibfnamefont {M.}~\bibnamefont
  {Rupp}}, \bibinfo {author} {\bibfnamefont {A.}~\bibnamefont {Tkatchenko}},
  \bibinfo {author} {\bibfnamefont {K.-R.}\ \bibnamefont {M\"uller}}, \ and\
  \bibinfo {author} {\bibfnamefont {O.~A.}\ \bibnamefont {von Lilienfeld}},\
  }\href@noop {} {\bibfield  {journal} {\bibinfo  {journal} {Phys. Rev. Lett.}\
  }\textbf {\bibinfo {volume} {108}},\ \bibinfo {pages} {058301} (\bibinfo
  {year} {2012})}\BibitemShut {NoStop}%
\bibitem [{\citenamefont {Huo}\ and\ \citenamefont
  {Rupp}(2017)}]{Huo2017arxiv}%
  \BibitemOpen
  \bibfield  {author} {\bibinfo {author} {\bibfnamefont {H.}~\bibnamefont
  {Huo}}\ and\ \bibinfo {author} {\bibfnamefont {M.}~\bibnamefont {Rupp}},\
  }\href@noop {} {\bibfield  {journal} {\bibinfo  {journal} {arXiv:1704.06439}\
  } (\bibinfo {year} {2017})}\BibitemShut {NoStop}%
\bibitem [{\citenamefont {Hirn}\ \emph {et~al.}(2017)\citenamefont {Hirn},
  \citenamefont {Mallat},\ and\ \citenamefont {Poilvert}}]{Hirn2017}%
  \BibitemOpen
  \bibfield  {author} {\bibinfo {author} {\bibfnamefont {M.}~\bibnamefont
  {Hirn}}, \bibinfo {author} {\bibfnamefont {S.}~\bibnamefont {Mallat}}, \ and\
  \bibinfo {author} {\bibfnamefont {N.}~\bibnamefont {Poilvert}},\ }\href@noop
  {} {\bibfield  {journal} {\bibinfo  {journal} {Multiscale Modeling \&
  Simulation}\ }\textbf {\bibinfo {volume} {15}},\ \bibinfo {pages} {827}
  (\bibinfo {year} {2017})}\BibitemShut {NoStop}%
\bibitem [{\citenamefont {Willatt}\ \emph {et~al.}(2019)\citenamefont
  {Willatt}, \citenamefont {Musil},\ and\ \citenamefont
  {Ceriotti}}]{willatt2019}%
  \BibitemOpen
  \bibfield  {author} {\bibinfo {author} {\bibfnamefont {M.~J.}\ \bibnamefont
  {Willatt}}, \bibinfo {author} {\bibfnamefont {F.}~\bibnamefont {Musil}}, \
  and\ \bibinfo {author} {\bibfnamefont {M.}~\bibnamefont {Ceriotti}},\
  }\href@noop {} {\bibfield  {journal} {\bibinfo  {journal} {The Journal of
  Chemical Physics}\ }\textbf {\bibinfo {volume} {150}},\ \bibinfo {pages}
  {154110} (\bibinfo {year} {2019})}\BibitemShut {NoStop}%
\bibitem [{Note1()}]{Note1}%
  \BibitemOpen
  \bibinfo {note} {Evaluation of the integral for $p>1$ require some form of
  regularization or short-distance cutoff to remove the singularity for
  $\protect \mathbf {r}\rightarrow \protect \mathbf {r}_i$}\BibitemShut
  {NoStop}%
\bibitem [{\citenamefont {Dreizler}\ and\ \citenamefont
  {Gross}(2012)}]{Dreizler2012}%
  \BibitemOpen
  \bibfield  {author} {\bibinfo {author} {\bibfnamefont {R.}~\bibnamefont
  {Dreizler}}\ and\ \bibinfo {author} {\bibfnamefont {E.}~\bibnamefont
  {Gross}},\ }\href@noop {} {\emph {\bibinfo {title} {Density Functional
  Theory: An Approach to the Quantum Many-Body Problem}}}\ (\bibinfo
  {publisher} {Springer Berlin Heidelberg},\ \bibinfo {year}
  {2012})\BibitemShut {NoStop}%
\bibitem [{\citenamefont {Huang}\ and\ \citenamefont {von
  Lilienfeld}(2017)}]{Huang2019arxiv}%
  \BibitemOpen
  \bibfield  {author} {\bibinfo {author} {\bibfnamefont {B.}~\bibnamefont
  {Huang}}\ and\ \bibinfo {author} {\bibfnamefont {O.~A.}\ \bibnamefont {von
  Lilienfeld}},\ }\href@noop {} {\bibfield  {journal} {\bibinfo  {journal}
  {arXiv:1707.04146}\ } (\bibinfo {year} {2017})}\BibitemShut {NoStop}%
\bibitem [{\citenamefont {De}\ \emph {et~al.}(2016)\citenamefont {De},
  \citenamefont {Bart{\'{o}}k}, \citenamefont {Cs{\'{a}}nyi},\ and\
  \citenamefont {Ceriotti}}]{de+16pccp}%
  \BibitemOpen
  \bibfield  {author} {\bibinfo {author} {\bibfnamefont {S.}~\bibnamefont
  {De}}, \bibinfo {author} {\bibfnamefont {A.~A.~P.}\ \bibnamefont
  {Bart{\'{o}}k}}, \bibinfo {author} {\bibfnamefont {G.}~\bibnamefont
  {Cs{\'{a}}nyi}}, \ and\ \bibinfo {author} {\bibfnamefont {M.}~\bibnamefont
  {Ceriotti}},\ }\href {\doibase 10.1039/C6CP00415F} {\bibfield  {journal}
  {\bibinfo  {journal} {Phys. Chem. Chem. Phys.}\ }\textbf {\bibinfo {volume}
  {18}},\ \bibinfo {pages} {13754} (\bibinfo {year} {2016})}\BibitemShut
  {NoStop}%
\bibitem [{\citenamefont {Grisafi}\ \emph {et~al.}(2019)\citenamefont
  {Grisafi}, \citenamefont {Wilkins}, \citenamefont {Willatt},\ and\
  \citenamefont {Ceriotti}}]{grisafi2019-arxiv}%
  \BibitemOpen
  \bibfield  {author} {\bibinfo {author} {\bibfnamefont {A.}~\bibnamefont
  {Grisafi}}, \bibinfo {author} {\bibfnamefont {D.~M.}\ \bibnamefont
  {Wilkins}}, \bibinfo {author} {\bibfnamefont {M.~J.}\ \bibnamefont
  {Willatt}}, \ and\ \bibinfo {author} {\bibfnamefont {M.}~\bibnamefont
  {Ceriotti}},\ }\href@noop {} {\bibfield  {journal} {\bibinfo  {journal}
  {arXiv:1904.01623}\ } (\bibinfo {year} {2019})}\BibitemShut {NoStop}%
\bibitem [{\citenamefont {Ewald}(1921)}]{Ewald1921}%
  \BibitemOpen
  \bibfield  {author} {\bibinfo {author} {\bibfnamefont {P.~P.}\ \bibnamefont
  {Ewald}},\ }\href@noop {} {\bibfield  {journal} {\bibinfo  {journal} {Annalen
  der Physik}\ }\textbf {\bibinfo {volume} {369}},\ \bibinfo {pages} {253}
  (\bibinfo {year} {1921})}\BibitemShut {NoStop}%
\bibitem [{\citenamefont {Essmann}\ \emph {et~al.}(1995)\citenamefont
  {Essmann}, \citenamefont {Perera}, \citenamefont {Berkowitz}, \citenamefont
  {Darden}, \citenamefont {Lee},\ and\ \citenamefont {Pedersen}}]{Essmann1995}%
  \BibitemOpen
  \bibfield  {author} {\bibinfo {author} {\bibfnamefont {U.}~\bibnamefont
  {Essmann}}, \bibinfo {author} {\bibfnamefont {L.}~\bibnamefont {Perera}},
  \bibinfo {author} {\bibfnamefont {M.~L.}\ \bibnamefont {Berkowitz}}, \bibinfo
  {author} {\bibfnamefont {T.}~\bibnamefont {Darden}}, \bibinfo {author}
  {\bibfnamefont {H.}~\bibnamefont {Lee}}, \ and\ \bibinfo {author}
  {\bibfnamefont {L.~G.}\ \bibnamefont {Pedersen}},\ }\href@noop {} {\bibfield
  {journal} {\bibinfo  {journal} {The Journal of Chemical Physics}\ }\textbf
  {\bibinfo {volume} {103}},\ \bibinfo {pages} {8577} (\bibinfo {year}
  {1995})}\BibitemShut {NoStop}%
\bibitem [{\citenamefont {Cahill}(2013)}]{Cahill2013}%
  \BibitemOpen
  \bibfield  {author} {\bibinfo {author} {\bibfnamefont {K.}~\bibnamefont
  {Cahill}},\ }\href@noop {} {\emph {\bibinfo {title} {Physical Mathematics}}}\
  (\bibinfo  {publisher} {Cambridge University Press},\ \bibinfo {year}
  {2013})\BibitemShut {NoStop}%
\bibitem [{\citenamefont {Allen}\ and\ \citenamefont
  {Tildesley}(1989)}]{Allen1989}%
  \BibitemOpen
  \bibfield  {author} {\bibinfo {author} {\bibfnamefont {M.~P.}\ \bibnamefont
  {Allen}}\ and\ \bibinfo {author} {\bibfnamefont {D.~J.}\ \bibnamefont
  {Tildesley}},\ }\href@noop {} {\emph {\bibinfo {title} {Computer Simulation
  of Liquids}}}\ (\bibinfo  {publisher} {Clarendon Press},\ \bibinfo {year}
  {1989})\BibitemShut {NoStop}%
\bibitem [{\citenamefont {Plimpton}(1995)}]{Plimpton1995}%
  \BibitemOpen
  \bibfield  {author} {\bibinfo {author} {\bibfnamefont {S.}~\bibnamefont
  {Plimpton}},\ }\href@noop {} {\bibfield  {journal} {\bibinfo  {journal} {J.
  Comp. Phys.}\ }\textbf {\bibinfo {volume} {117}},\ \bibinfo {pages} {1}
  (\bibinfo {year} {1995})}\BibitemShut {NoStop}%
\bibitem [{\citenamefont {Burns}\ \emph {et~al.}(2017)\citenamefont {Burns},
  \citenamefont {Faver}, \citenamefont {Zheng}, \citenamefont {Marshall},
  \citenamefont {Smith}, \citenamefont {Vanommeslaeghe}, \citenamefont
  {MacKerell}, \citenamefont {Merz},\ and\ \citenamefont
  {Sherrill}}]{Burns2017}%
  \BibitemOpen
  \bibfield  {author} {\bibinfo {author} {\bibfnamefont {L.~A.}\ \bibnamefont
  {Burns}}, \bibinfo {author} {\bibfnamefont {J.~C.}\ \bibnamefont {Faver}},
  \bibinfo {author} {\bibfnamefont {Z.}~\bibnamefont {Zheng}}, \bibinfo
  {author} {\bibfnamefont {M.~S.}\ \bibnamefont {Marshall}}, \bibinfo {author}
  {\bibfnamefont {D.~G.}\ \bibnamefont {Smith}}, \bibinfo {author}
  {\bibfnamefont {K.}~\bibnamefont {Vanommeslaeghe}}, \bibinfo {author}
  {\bibfnamefont {A.~D.}\ \bibnamefont {MacKerell}}, \bibinfo {author}
  {\bibfnamefont {K.~M.}\ \bibnamefont {Merz}}, \ and\ \bibinfo {author}
  {\bibfnamefont {C.~D.}\ \bibnamefont {Sherrill}},\ }\href {\doibase
  10.1063/1.5001028} {\bibfield  {journal} {\bibinfo  {journal} {J. Chem.
  Phys.}\ }\textbf {\bibinfo {volume} {147}},\ \bibinfo {pages} {161727}
  (\bibinfo {year} {2017})}\BibitemShut {NoStop}%
\bibitem [{\citenamefont {Blum}\ \emph {et~al.}(2009)\citenamefont {Blum},
  \citenamefont {Gehrke}, \citenamefont {Hanke}, \citenamefont {Havu},
  \citenamefont {Havu}, \citenamefont {Ren}, \citenamefont {Reuter},\ and\
  \citenamefont {Scheffler}}]{Blum2009}%
  \BibitemOpen
  \bibfield  {author} {\bibinfo {author} {\bibfnamefont {V.}~\bibnamefont
  {Blum}}, \bibinfo {author} {\bibfnamefont {R.}~\bibnamefont {Gehrke}},
  \bibinfo {author} {\bibfnamefont {F.}~\bibnamefont {Hanke}}, \bibinfo
  {author} {\bibfnamefont {P.}~\bibnamefont {Havu}}, \bibinfo {author}
  {\bibfnamefont {V.}~\bibnamefont {Havu}}, \bibinfo {author} {\bibfnamefont
  {X.}~\bibnamefont {Ren}}, \bibinfo {author} {\bibfnamefont {K.}~\bibnamefont
  {Reuter}}, \ and\ \bibinfo {author} {\bibfnamefont {M.}~\bibnamefont
  {Scheffler}},\ }\href@noop {} {\bibfield  {journal} {\bibinfo  {journal}
  {Computer Physics Communications}\ }\textbf {\bibinfo {volume} {180}},\
  \bibinfo {pages} {2175 } (\bibinfo {year} {2009})}\BibitemShut {NoStop}%
\bibitem [{\citenamefont {Glielmo}\ \emph {et~al.}(2018)\citenamefont
  {Glielmo}, \citenamefont {Zeni},\ and\ \citenamefont
  {De~Vita}}]{Glielmo2018}%
  \BibitemOpen
  \bibfield  {author} {\bibinfo {author} {\bibfnamefont {A.}~\bibnamefont
  {Glielmo}}, \bibinfo {author} {\bibfnamefont {C.}~\bibnamefont {Zeni}}, \
  and\ \bibinfo {author} {\bibfnamefont {A.}~\bibnamefont {De~Vita}},\
  }\href@noop {} {\bibfield  {journal} {\bibinfo  {journal} {Phys. Rev. B}\
  }\textbf {\bibinfo {volume} {97}},\ \bibinfo {pages} {184307} (\bibinfo
  {year} {2018})}\BibitemShut {NoStop}%
\bibitem [{\citenamefont {Drautz}(2019)}]{Drautz2019}%
  \BibitemOpen
  \bibfield  {author} {\bibinfo {author} {\bibfnamefont {R.}~\bibnamefont
  {Drautz}},\ }\href@noop {} {\bibfield  {journal} {\bibinfo  {journal} {Phys.
  Rev. B}\ }\textbf {\bibinfo {volume} {99}},\ \bibinfo {pages} {014104}
  (\bibinfo {year} {2019})}\BibitemShut {NoStop}%
\end{thebibliography}
\end{document}